# Coulomb-driven broken-symmetry states in doubly gated suspended bilayer graphene

R. T. Weitz, M. T. Allen, B. E. Feldman, J. Martin and A. Yacoby*

*Department of Physics, Harvard University, 11 Oxford St., Cambridge, 02138 MA, U.S.A.*
*yacoby@physics.harvard.edu*

*One sentence summary:* Doubly suspended bilayer graphene reveals Coulomb driven broken symmetry states at large *B* as well as at *B = 0*.

**Abstract**

The non-interacting energy spectrum of graphene and its bilayer counterpart consists of multiple degeneracies owing to the inherent spin, valley and layer symmetries. Interactions among charge carriers are expected to spontaneously break these symmetries, leading to gapped ordered states. In the quantum Hall regime these states are predicted to be ferromagnetic in nature whereby the system becomes spin polarized, layer polarized or both. In bilayer graphene, due to its parabolic dispersion, interaction-induced symmetry breaking is already expected at zero magnetic field. In this work, the underlying order of the various broken-symmetry states is investigated in bilayer graphene that is suspended between top and bottom gate electrodes. By controllably breaking the spin and sublattice symmetries we are able to deduce the order parameter of the various quantum Hall ferromagnetic states. At small carrier densities, we identify for the first time three distinct broken symmetry states, one of which is consistent with either spontaneously broken time-reversal symmetry or spontaneously broken rotational symmetry.

Due to their unique electronic properties, mono- and bilayer graphene have attracted significant research efforts since their initial experimental discoveries (*1-6*). In bilayer graphene, the application of a perpendicular electric field allows for the continuous opening of a band gap in the density of states (*7-14*), a unique property that renders it a promising candidate for applications in electronics and optoelectronics (*5*). Perhaps more exciting in terms of basic science are the discoveries of an unconventional quantum Hall effect in mono- and bilayer graphene, which arise from the chiral nature of the charge carriers in these materials (*2-6, 15*). In monolayers, the sequence of Hall plateaus is shifted by a half integer and each Landau level (LL) is 4-fold degenerate due to spin and valley degrees of freedom. In bilayer graphene, an even richer picture emerges in the lowest LL due to an additional degeneracy between the zeroth and first orbital LLs, giving rise to an 8-fold degeneracy. Systems in which multiple LLs are degenerate have been shown to give rise to broken-symmetry states caused by electron-electron interactions (*16*). Such interaction-induced broken-symmetry states in the lowest LL of bilayer graphene have been theoretically predicted (*17, 18*) and experimentally observed (*19-23*), but the nature of their order parameters is still debated. For example, two possible order parameters that have been suggested for the gapped phase at $\nu = 0$ at large magnetic fields are either layer or spin polarization (*18*). Moreover, recent theoretical works predict that even in the absence of external magnetic and electric fields the parabolic dispersion of bilayer graphene can lead to broken-symmetry states that are induced by interactions among the charge carriers (*24-31*). In this work, we map out the various broken symmetry states as a function of external magnetic and electric field. The nature of these phases can be deduced by investigating their stability under the variation of these symmetry-breaking fields.

Broken symmetry states driven by effects of interaction rely critically on high sample quality. We have therefore developed a method to fabricate samples in which bilayer graphene is suspended between a top gate electrode and the substrate. This allows us to combine the high quality of suspended devices with the ability to independently control electron density and perpendicular electric field $E_\perp$.

A false color scanning electron microscope (SEM) image of a typical device is shown in **Figure 1a**. The suspended graphene (red) is supported by gold contacts (yellow). Suspended top gates (blue) can be designed to cover only part of the graphene (left device) or to fully overlap the entire flake (right device), including part of the contacts. We have investigated both types of devices, which show very similar characteristics. The fabrication of such two-terminal devices (detailed in the Supporting Online Material) is schematically illustrated in **Figure 1a**. In order to improve sample quality, we current anneal (*32, 33*) our devices in vacuum at 4 K prior to measurement.

A clear indication of the high quality of our suspended flakes can be seen from the dependence of resistance versus applied electric field at zero magnetic field. As theoretically predicted (*7, 9, 34*) and experimentally verified in transport (*11, 13*) and optical (*12-14*) experiments, an applied perpendicular electric field $E_\perp$ opens a gap $\Delta \sim E_\perp/(kd)$ in the otherwise gapless dispersion of bilayer graphene, as schematically shown in the lower left inset to **Figure 1b**. Here, $d$ is the distance between the graphene sheets and $k$ is a constant that accounts for imperfect screening of the external electric field by the bilayer (*7, 10, 28*). The conductance of our suspended bilayer graphene sheet at 100 mK as a function of back ($V_b$) and top gate ($V_t$) voltage is shown in **Figure 1b**. The opening of a field-induced band gap is apparent from the decreased conductance at the charge neutrality point with increasing applied electric field $E_\perp = (\alpha V_t - \beta V_b)/2e\varepsilon_0$. Here, $\alpha$ and $\beta$ are the gate coupling factors for $V_t$ and $V_b$ respectively, which we determine from the Landau fan in the quantum Hall regime (see Supporting Online Material), $e$ the electron charge and $\varepsilon_0$ the vacuum permittivity. The presence of an energy gap is also illustrated by line cuts in **Figure 1c,** which show the resistance at constant $E_\perp$ as function of total density $n = (\alpha V_t + \beta V_b)$. Here it can be seen that by varying the density, the conductance at the highest electric field can be changed by a factor $10^3$. Line traces of the maximum sheet resistance as a function of electric field at various temperatures are shown in **Figure 1d**, in which an exponential dependence of the resistance on electric field, as well as decrease in resistance with temperature are clearly seen. The sheet resistance at 100 mK increases by more than a factor of $2 \times 10^3$ (from 8 kΩ/square to *20* M Ω/square) between $E_\perp = 20$ mV/nm and 90 mV/nm. Compared to previous measurements of dually gated graphene bilayers embedded in a dielectric (*11*), our measurements show an increase by a factor of $10^3$ in resistivity at the same electric field, clearly attesting to the high quality of our flakes. The unexpected finding of a non-monotonous dependence of the resistance at small applied electric field will be discussed towards the end of the manuscript. The oscillations in the conductance traces shown in **Figure 1c** are repeatable and result from mesoscopic conductance fluctuations. A detailed description of the temperature dependence and these fluctuations are presented in the Supporting Online Material.

We now describe the conductance of our samples at nonzero magnetic field, focusing in particular on the evolution of different LLs in an applied electric field. **Figure 2a-d** shows the two-terminal conductance as a function of density and electric field at various different magnetic fields. As was previously reported, the 8-fold degeneracy of the lowest LL is fully lifted due to electron-electron interactions (*19, 20, 23*), and plateaus at filling factors $\nu = 0, \pm1, \pm2, \pm3$ can be identified by their slope in a fan diagram. Vertical line cuts that correspond to a constant filling factor show that the conductance is quantized except at particular values of electric field, denoted by stars or dots in **Figures 2 a-d.** The $\nu = 0$ state is quantized except near two values of the applied electric field (marked with dots). This value of electric field increases as B increases. The $\nu = \pm2$ state is quantized for all electric fields except near $E_\perp = 0$, finally $\nu = \pm4$ is quantized for all electric fields.

A qualitative understanding of this phenomenology can be gained from the simplified scheme of the LL energies as a function of electric field shown in **Figure 2e** (gray lines) for nonzero magnetic field**.** We start by neglecting the breaking of the orbital degree of freedom so that the LLs are each doubly degenerate (*15*). Let us assume, in accordance with theoretical predictions, that at $E_\perp = 0$ only the spin degeneracy is lifted, giving rise to a spin polarized $\nu = 0$ quantum Hall ferromagnetic state (*18, 28*). In the lowest LL, the layer and valley index are equivalent (*15*) so that an electric field which favors one of the layers controls directly the valley-pseudospin. As a result, LLs of quantum number U (upper layer) or L (lower layer) are expected to have different slopes in electric field. At several points, marked by dots or stars in the schematic of **Figure 2e**, LL crossings occur. A single LL crossing is seen at $E_\perp = 0$ each for $\nu = 2$ and $\nu = -2$ and two LL crossings are seen for $\nu = 0$ at nonzero $E_\perp$. We hypothesize that these crossings are responsible for the increased conductance in our transport experiments (*35*). It is worthwhile to note that **Figure 2e** shows LL energy whereas we have direct control over the density of the bilayer rather than its chemical potential. However, on a quantum Hall plateau, the chemical potential lies between two LL energies, which enables us to relate our scheme in **Figure 2e** to our transport data as detailed below (see also Supporting Online Material).

The LL crossings at zero average carrier density are marked by dots in **Figure 2**. In our proposed scheme, these transitions separate a spin polarized $\nu = 0$ state at low electric fields (I) from two layer polarized $\nu = 0$ states at large electric fields (II) of opposite layer polarization. The line cut in **Figure 2c** at constant filling factor shows that insulating $\nu = 0$ states are well-developed at zero electric field and at large electric fields, but the crossover between these states is marked by a region of increased conductance. We would like to point out that the experimentally observed large resistance of the phase at $E_\perp = 0$ (*19, 20*) is at variance with the theoretical prediction of percolating edge modes (*36*) in the case of a spin polarized $\nu = 0$ state. Possible reasons include mixing of counter-propagating edge modes and subsequent opening of gaps (*37*).

The LL crossings at $\nu = \pm 2$ near zero electric field are marked by stars in **Figure 2**. An example of such a crossing is apparent in the vertical line cut shown in **Figure 2b**. There the conductance at $\nu = \pm 2$ increases near zero electric field, which is explained in our scheme by a crossing between two LLs of identical spin polarization but opposite layer polarization. The consequence of these crossings is that only $\nu = 0$ and $\nu = \pm 4$ states occur at zero electric field, whereas layer polarized $\nu = \pm 2$ states emerge only at finite electric field, as is apparent in the horizontal line cuts of **Figure 2b**.

Assuming that the $\nu = \pm 1$ and $\pm 3$ states are partially layer polarized (*18*), the electric field should also induce a splitting at these filling factors (inset to **Figure 2e**) that is however more fragile and therefore only seen at larger magnetic fields. Our simple model predicts that the $\nu = \pm 3$ state will have a crossing at zero electric field and we attribute the increase in conductance at $E_\perp = 0$ in the left line cut in **Figure 2d** to be a signature of this behavior. A line cut at $\nu = \pm 3$ at even higher magnetic field showing a better quantization is shown in the Supporting Online Material. The $\nu = \pm 1$ state is expected to have three crossings: one at zero electric field and two near the same nonzero electric field at which the transition between the $\nu = 0$ states is observed. All three of these crossings are apparent from the regions of increased conductance in the right line cut in **Figure 2d**. The observation that $\nu = \pm 1$ and $\pm 3$ states are enhanced with electric field is an indication of their partial layer polarization. Theory nonetheless predicts, that $\nu = \pm 1, \pm 2, \pm 3$ should also be seen at $E_\perp = 0$ (*18*). However, the predicted energy gaps of these states are expected to be considerably smaller than that of $\nu = 0$ and hence visible only at large magnetic field. Indeed, at $B > 4$ T these broken symmetry states at $E_\perp = 0$ can also be observed in our data (see Supporting Online Material).

We now discuss the various phases near zero carrier density in more detail. **Figure 3a** shows the conductance at the charge neutrality point as a function of electric and magnetic field. The regions of high conductance mark the transition between the low- and high-field $\nu = 0$ states (highlighted with dots in **Figure 2**). It can be seen that the transition moves out to larger electric fields as the magnetic field is increased which implies that phase I is stable at large magnetic fields and is destabilized by an electric field. This observation is also consistent with our energy diagram (**Figure 2e**), which predicts the $\nu = 0$ LL crossing to move out to higher electric fields upon increase in magnetic field. The fact that the stability of this phase increases with magnetic field reconfirms our initial assumption that phase I is spin polarized. In contrast, the phase at large electric fields (phase II) is stabilized by an electric field, consistent with it being layer polarized.

The dependence of the transition region between the two phases on electric and magnetic field is shown more clearly in **Figure 3b**, where we have extracted the maximum conductance in **Figure 3a** for all positive values of electric field. At large magnetic field, the transition line is linear, independent of sample quality and temperature (we have investigated a total of 5 samples at temperatures between 100 mK and 5 K). At magnetic fields below about 2 T, however, the transition line is only linear in the highest quality sample and at low temperatures (inset to **Figure 3a** and dashed line in **Figure 3b**). A surprising observation is that the extrapolation of the linear dependence from high B down to B = 0 results in a crossing at a non-zero electric field $E_{off}$. Moreover the slope and $E_{off}$ that characterize the large B behavior seem to have no systematic dependence on sample quality and temperature T (**Figure 3c** and **d**). The conductance at the transition between the $\nu = 0$ phases decreases both with increasing B as well as with decreasing T (see Supporting Online Material). Such behavior is qualitatively expected given that LL mixing at these crossing points can lead to the opening of gaps in the spectrum.

To shed further light on the nature of the transition it is instructive to investigate the role of the Zeeman energy $E_z$ across the $\nu = 0$ transition in detail. A convenient method to do so is to tilt the sample with respect to the magnetic field thus altering $E_z$ and leaving all interaction dependent energy scales the same. **Figure 3b** shows both measurements under titled magnetic field (green line) and data taken in purely perpendicular field (blue line). It can be seen that the slope of the two transition lines are very similar which indicates that $E_z$ is negligible and the transition predominantly depends on the perpendicular component of the magnetic field indicating that exchange effects and the LL degeneracies play an important role. This explanation however does not provide an explanation for the B = 0 offset.

The crossover between two phases with the application of an electric field has also been predicted theoretically by Gorbar *et al.* (*38*) and Nandikishore *et al.* (*28*). Gorbar *et al.* (*38*) predict a transition from a quantum Hall ferromagnetic state at low electric fields to a insulating state dominated by magnetic catalysis with a slope of about $2\ mV/(nm*T)$. Nandkishore *et al.* (*28*) predict a transition from a quantum Hall ferromagnetic state to a layer insulating state at large electric fields, with a slope of $34\ mV/(nm*T)$. Our measured slope is about $11\ mV/(nm*T)$. A qualitative comparison between the experimentally obtained values and the theory is difficult because of the lack of knowledge of the screening that the bilayer provides to the applied electric field once LLs are formed.

A striking feature seen in the inset to **Figure 3a** and **Figure 3b** is that the transition line appears to have a non-zero offset in electric field, $E_{off} \approx 20\ mV/nm$. The $B = 0$ offset coincides with the extrapolated one from high fields when the sample quality is high and at low temperatures. This suggests that the transition from a spin to a layer polarized phase persists all the way down to $B = 0$. Careful measurements near $B = 0$ and $E_\perp = 0$ (**Figure 4a**) however show that this conclusion is incorrect. We will discuss possible origins for $E_{off}$ in the remainder of this manuscript.

The conductance close to the charge neutrality point at small electric and magnetic fields is shown in **Figure 4a**. The conductance at $B = 0$ and $E_\perp = 0$ exhibits a local minimum and increases upon increase of $E_\perp$ and $B$. The electric field value at which the conductance reaches is maximum coincides with $E_{off}$. Together with observation of a maximum of the conductance at a finite value $B_{off} = \pm 50$ $mT$ our observations suggest that neither the spin polarized phase I nor the layer polarized phase II extend down to $B = 0$ and $E_\perp = 0$. Instead, our measurements are consistent with the presence of a third phase at small electric and magnetic fields. The electric field dependence of the resistance is at odds with simple calculations of the band structure of bilayer graphene. In a tight binding model, bilayer graphene is expected to evolve from a gapless semimetal to a gapped semiconductor whose gap magnitude depends monotonically on electric field (*7*). However, our measurements show that the conductance does not decrease monotonically with electric field, instead exhibiting a local maximum at about ±20 mV/nm (**Figure 4b**). The maximum resistance at $B = 0$ and close to zero carrier density as a function of $E_\perp$ for different temperatures increases as T decreases (see **Figure 1d**) reaching *20 kΩ* at the lowest temperature of 100 mK. It is therefore strongly suggestive that the neutral bilayer graphene system is already gapped at $E_\perp = 0$ and B = 0 and that fields larger than $E_{off}$ or $B_{off}$ transfer the system into the layer or spin polarized phases (**Figure 4a**). In **Figure 4b** the dependence of the conductance as a function of density and electric field at B = 0 is shown. Also in this case a minimum of the conductance at small electric field and density can be discerned, indicative that the spontaneous phase is unstable away from the charge neutrality point. We note that the temperature dependence at the crossover field of $E_\perp = \pm 20$ mV/nm is weak (**Figure 1d**), consistent with the closure of a gap.

It has been pointed out theoretically (*24-31*) that electron-electron interactions can open a spontaneous gap in bilayer graphene at zero electric and magnetic fields, two of which fit well with our transport measurements. One such phase that is predicted to have a qualitatively similar phase diagram to the one we measure (*28*) is the anomalous quantum Hall insulating state that spontaneously breaks time reversal symmetry. This state is characterized by an antiferromagnetic ordering in the pseudospin (layer). In this phase electrons from one valley occupy one of the layers whereas electrons from the other valley occupy the other layer. An important feature of this state is that it supports topologically protected current carrying edge modes and is therefore predicted to have a finite conductance. This is consistent with our observations of a non-diverging resistance at B = 0 and $E_\perp = 0$ as T is lowered. Furthermore, our observation of a temperature dependence of the resistance in this phase is indicative of the presence of a gap. While our observed phase diagram agrees qualitatively with that predicted in (*28*) we find a quantitative disagreement between the measured and predicted E and B transition values. We find that the magnetic field at which the spontaneous phase breaks down is about 50 mT, an order of magnitude smaller than the theoretically predicted value of 500 mT (*28*). An applied electric field of about $E_{off}$ = 20 mV/nm quenches the spontaneous gap, compared to the predicted value of 26 mV/nm for the screened electric field. We however note that one needs to scale our measured $E_{off}$ by the screening constant *k* which means that our values differ by roughly a factor *k = 3* from the theory (*28*). A second possible scenario for the spontaneous order stems from the breaking of rotational symmetry in the presence of trigonal warping. This phase has been predicted to lead to a decreased density of states at the charge neutrality point due to the breaking of the system into two Dirac cones (*30*), consistent with our measured decrease in conductance at small densities. Further experimental support for the above two scenarios is given in (*39*).

**Acknowledgements**

The authors acknowledge discussions with R. Nandikshore, L. Levitov, M. Rudner, S. Sachdev, B. I. Halperin and G. Barak and experimental help by S. Foletti. RTW acknowledges financial support of the Alexander von Humboldt Foundation. This work was supported by the U.S. Department of Energy, Office of Basic Energy Sciences, Division of Materials Sciences and Engineering under Award #DE-SC0001819 (RTW, MTA, BEF), by the 2009 U.S. Office of Naval Research Multi University Research Initiative (MURI) on Graphene Advanced Terahertz Engineering (Gate) at MIT, Harvard and Boston University (JM) and by Harvard's NSEC under National Science Foundation award PHY-0646094. The authors declare to have no competing financial interests.

**Figures:**

**Figure 1** (a) False color scanning electron microscopy (SEM) image of a bilayer graphene flake suspended between Cr/Au electrodes with suspended Cr/Au top gates. A schematic of a cross section along a line marked by the two arrows is shown on the right including a brief depiction of the sample fabrication. We use a three step electron beam lithography process in which Cr/Au contacts are first fabricated on the bilayer graphene, followed by deposition of a $SiO_2$ spacer layer on top of the flake and finally the structure of a Cr/Au topgate above the $SiO_2$ layer. Subsequent etching of part of the $SiO_2$ leaves the graphene bilayer suspended between the top and bottom gate electrodes. (b) Conductance map as function of top and bottom gate voltage at T = 100 mK of a suspended bilayer graphene flake. (c) Line traces along the dashed lines in (b) showing the sheet resistance as function of charge carrier density at constant electric field. (d) Traces of the maximal sheet resistance at the charge neutrality point as function of applied electric field at different temperatures.

**Figure 2** (a)-(d) Maps of the conductance in units of $e^2/h$ as function of applied electric field $E_\perp$ and density $n$ at various constant magnetic fields at T = 60 mK. Line traces are taken from the data along the dashed lines. Horizontal line traces correspond to cuts at constant electric field, vertical line cuts correspond to cuts along constant filling factor. In (b) the orange line cut corresponds to the electric field dependence of $\nu = 0$ and the vertical line cuts to the density dependence of the conductance at two constant electric fields showing the emergence of $\nu = 2$ at large electric fields. In (c) the red vertical line cut corresponds to the electric field dependence of $\nu = 0$. The horizontal line cuts in green and blue show the density dependence of the conductance at two different electric fields showing the suppression of the $\nu = 0$ state at a finite electric field. In (d) the brown (violet) line cut shows the conductance as function of electric field at $\nu = 1$ ($\nu = 3$). (e) Schematic diagram of the energetic position of the lowest LL octet (grey lines) as function of $E_\perp$. The quantum numbers of the LLs are indicated in the boxes. The respective filling factors $\nu$ are indicted in the grey boxes. The LLs in the main scheme are all doubly degenerate in orbital quantum numbers. The impact of the electric field on the orbital LLs is shown in the two insets. The two different $\nu = 0$ states are marked with Roman numbers. In all images the crossing of LLs at zero electric field is marked with a star and the crossing at zero energy is marked with a dot.

**Figure 3** (a) Conductance in units of $e^2/h$ as function of applied electric and magnetic fields at average zero carrier density at 1.4 K. The transition between different $\nu = 0$ states is characterized by a region of increased conductance. Inset: Measurement taken at 100 mK in the case that the sample has been tilted with respect to the magnetic field plotted against the perpendicular component of the magnetic field in a different cooldown. The color scale of the inset reaches from 0 (blue) to 4.5 (red) $e^2/h$. (b) Comparison of the slope of the transition line in the case that the sample is perpendicular to the magnetic field and at a 45 degree angle. (c) and (d) comparison of the $E_{off}$ and high magnetic field slope of the transition line for different temperatures. Different colors correspond to different samples.

**Figure 4** Experimental evidence of a spontaneous gap in suspended bilayer graphene. **(a)** Detailed view of the conductivity at small electric and magnetic fields and average zero carrier density. We note that the color scale has been restricted between 5 and 6 $e^2/h$ to highlight the observed effect. **(b)** Conductivity as function of electric field and density at zero magnetic field. **(c)** Two linecuts showing the resistivity at E = 0 and $E_{off}$ are also shown. The scans in **(a)** and **(b)** were taken in different cool downs which is why minimal conductance at zero magnetic and electric field is different in both figures.

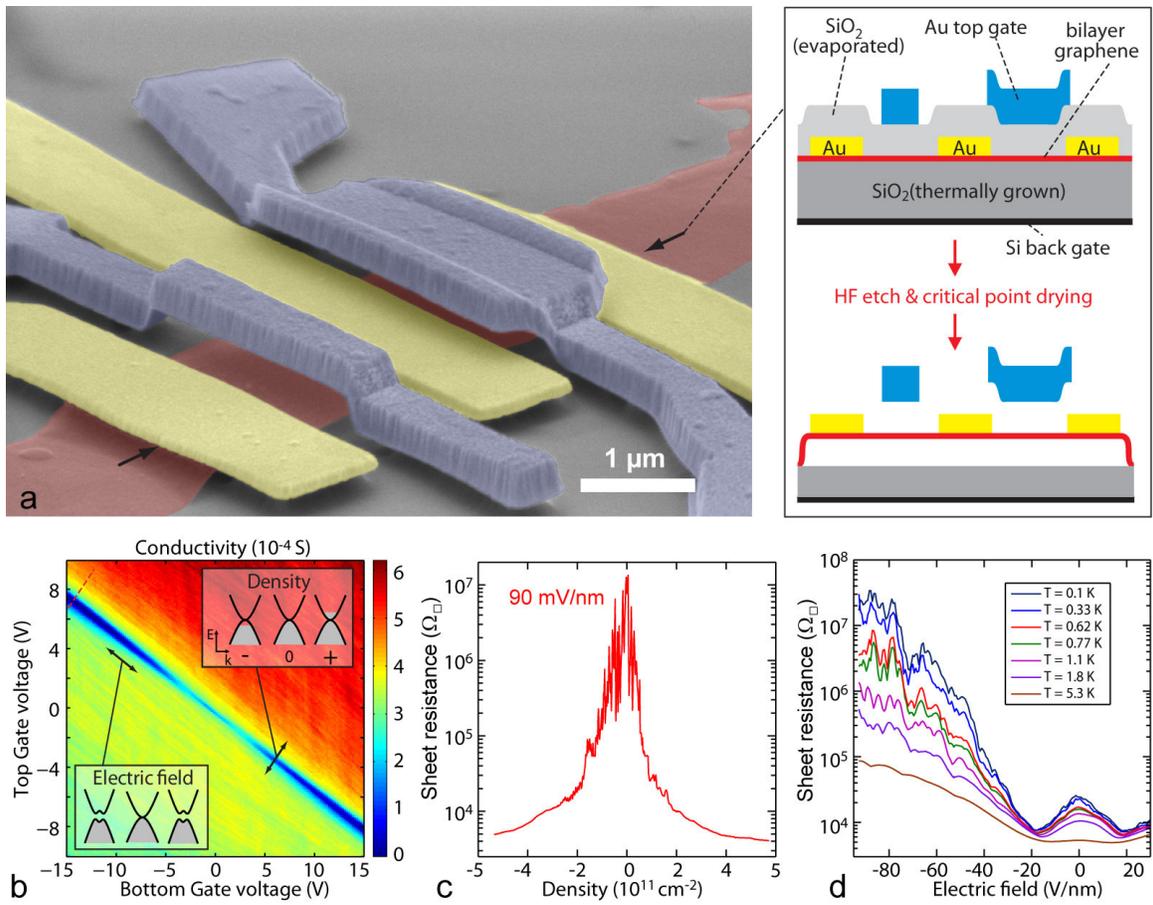

Figure 1

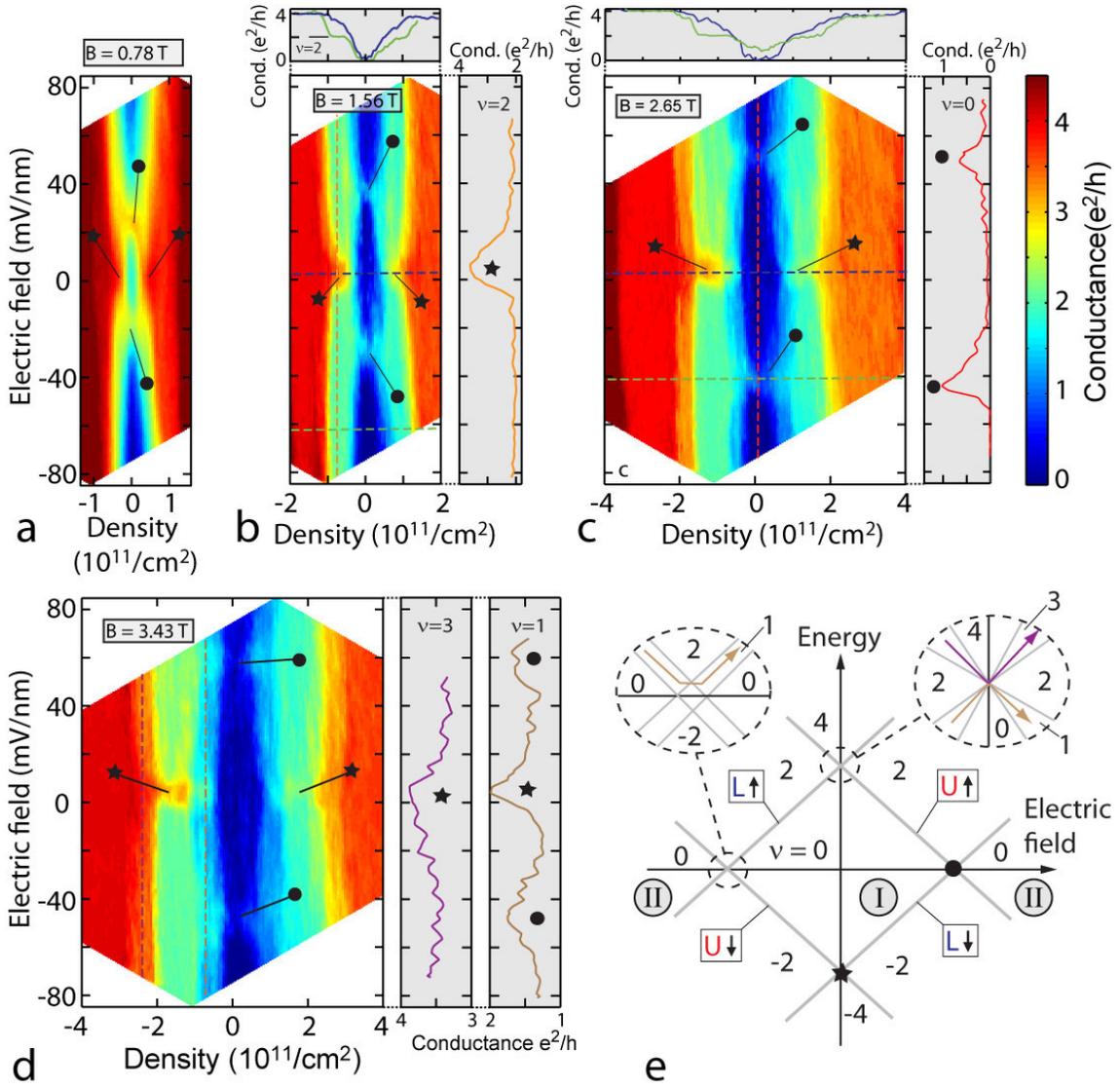

Figure 2

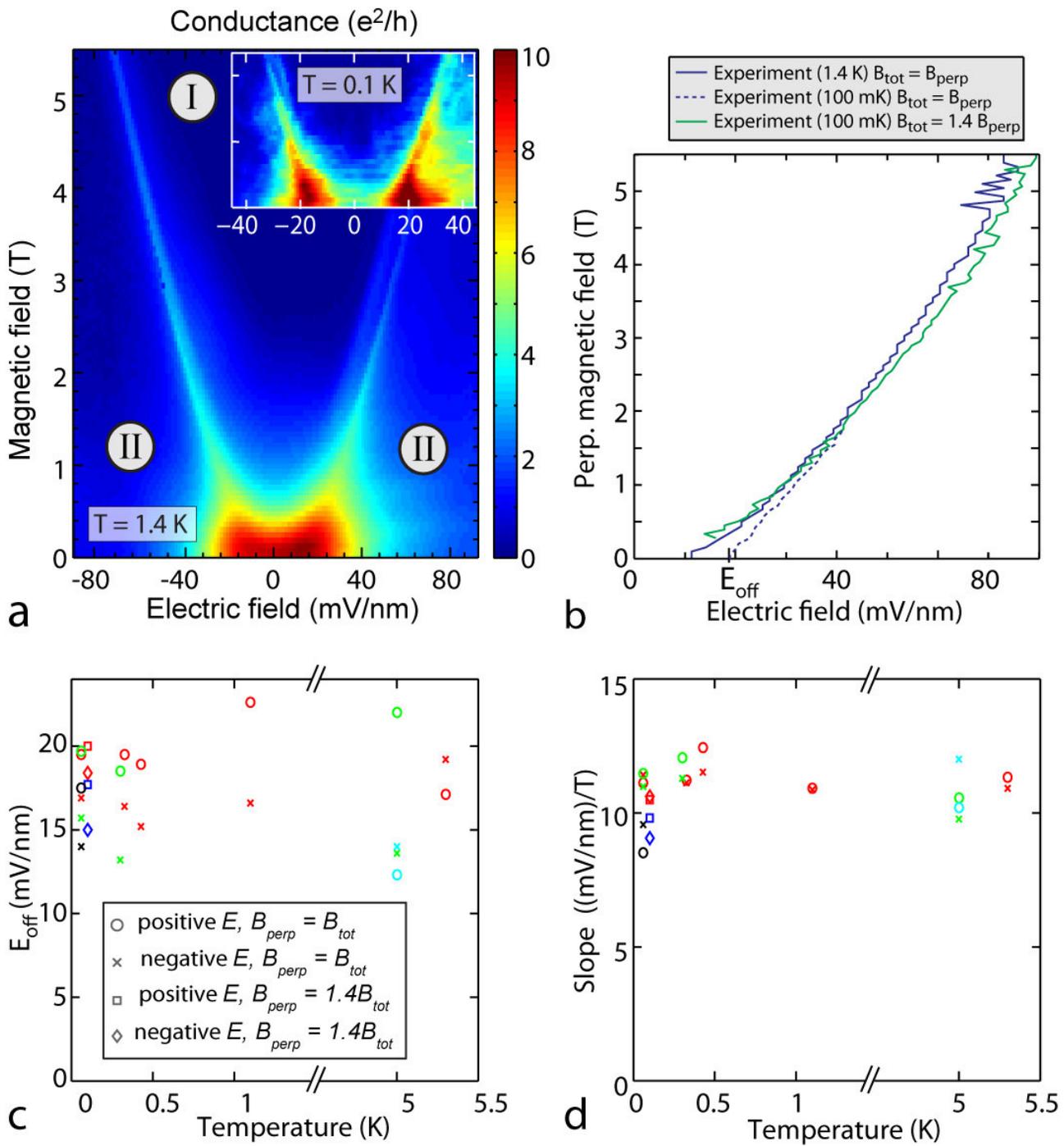

Figure 3

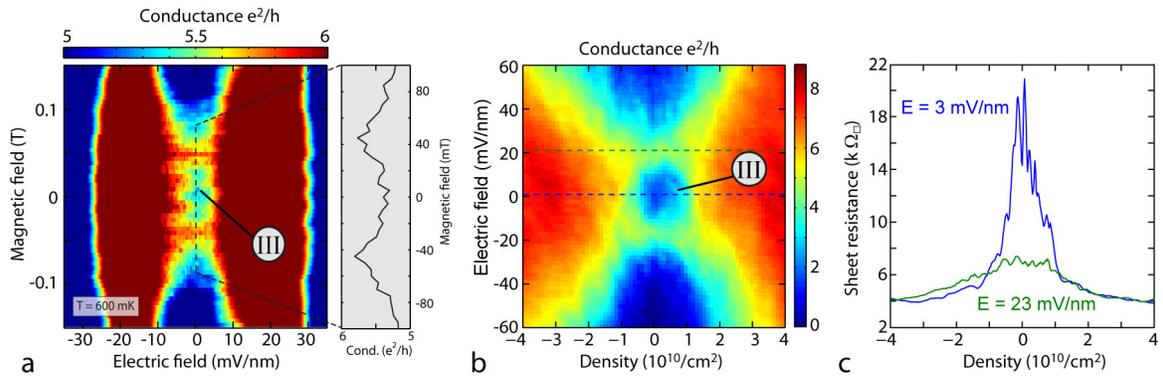

Figure 4

# Coulomb-driven broken-symmetry states in doubly gated suspended bilayer graphene

R. T. Weitz, M. T. Allen, B. E. Feldman, J. Martin and A. Yacoby*

*Department of Physics, Harvard University, 11 Oxford St., Cambridge, 02138 MA, U.S.A.*
*yacoby@physics.harvard.edu*

**Supporting Online Material**

1. Sample fabrication
2. Temperature dependence of the resistance at the charge neutrality point and B = 0
3. Mesoscopic conductance fluctuations at B = 0
4. Model of the dispersion of the LL energy in an electric field and its relation to density
5. Map of the conductance as function of electric field and density at B=7.8 T
6. Temperature dependence of the conductance at the transition between the two high magnetic field $\nu$ = 0 phases

1. Sample fabrication

Substrate cleaning and graphene deposition are performed as described by Feldman et al.(*1*). Graphene bilayers are identified by their contrast in an optical microscope and by their characteristic quantum Hall effect. Suitable flakes are contacted with Cr/Au contacts (3 nm/100 nm) by standard electron beam lithography, thermal metal evaporation and lift-off in acetone. Silicon dioxide is structured on top of graphene bilayers in a second electron beam lithography step, electron beam silicon dioxide evaporation and lift off. The silicon dioxide is used as a spacer layer to separate the top gate from the flake. The final electron beam lithography step is used to pattern top gates that are suspended from the substrate in the areas that the silicon oxide had been evaporated. Finally, the device is immersed into 5:1 buffered oxide etch for 90s and dried in methanol in a critical point dryer.

We have investigated a total of 5 samples. The data represented in **Figure 1**, the inset of **Figure 3a**, **Figure 4b**, **S1** and **S5** were taken in one of the samples in cool down 1. The data shown in **Figure 3**, **4a**, **S2** and **S4** were taken on the same sample in cool down 2. Data for **Figure 2** were taken from another sample. Data from all samples were used for **Figures 3c** and **3d**.

## 2. Temperature dependence of the sheet resistance at B=0

In transport experiments, the opening of a band gap $\Delta$ is manifested by a resistance that increases exponentially with gap size at zero average carrier density $R(T) = R_0 \exp(\Delta/2k_bT)$, where, $R_0$ is a constant, $k_b$ is Boltzmann's constant and $T$ is the temperature. At constant electric field $E_\perp$, the transport gap $E_{transport}=\Delta/2$ can be extracted from the linear fit of the resistance at the charge neutrality point to an Arrhenius plot, as shown in **Figure S1a**. The solid line is a linear fit which yields $E_{transport}$ = 0.31 meV disregarding the two lowest temperature data points. The saturation of the resistance below 600 mK may be due to tunneling between localized states. **Figure S1b** shows the maximum gap size obtained from Arrhenius plots at different densities near the charge neutrality point at a given electric field (again disregarding the lowest two temperatures in the fit). The gap size increases approximately linearly with the electric field, in agreement with theoretical predictions for a gap induced by an electric field (*2*).

Though the evolution of the gap with electric field agrees with theoretical predictions, the magnitudes

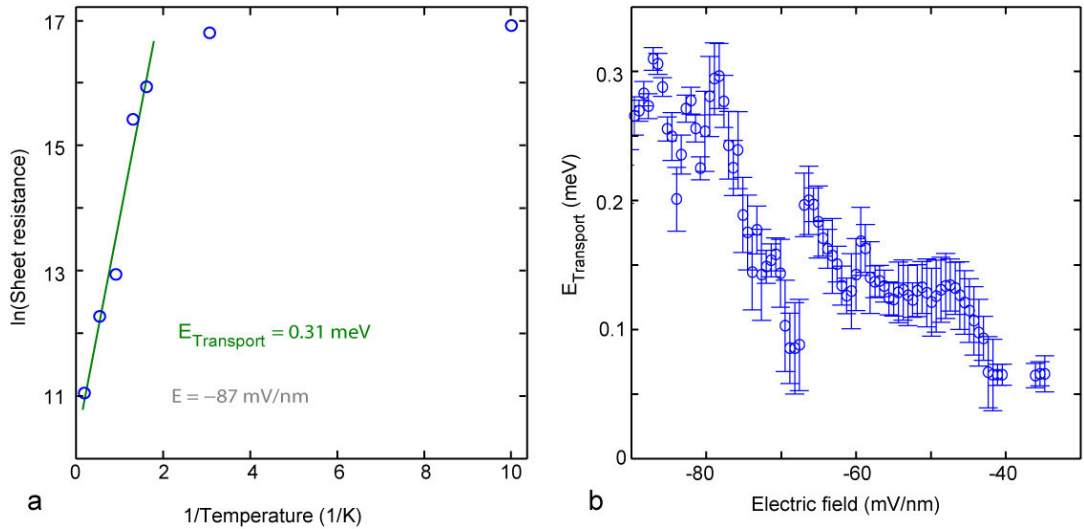

**Figure S1: (a)** Arrhenius plot of the resistance at a constant electric field of -87 mV/nm. **(b)** Dependence of the transport activation energy as function of electric field

of $E_{transport}$ are more than a factor 10 smaller than one would expect from the applied electric field. For example, at $E_\perp = -87\ mV/nm$ the expected activation gap would be $E_{theor}$ = 5 meV (*2*). Here, we again used a constant screening of *k = 3* (*2-4*). A possible explanation for this discrepancy is that even though disorder is small, it leads to a strong local variation of the Fermi level because for small overall carrier densities, the Fermi level has to shift by a significant amount in order to screen the charge disorder. Such a strong spatial variation could lead the discrepancy between the observed transport gap and the theoretically expected gap. A second explanation are the possible presence of states in the bandgap at the edge of the sample that have been predicted to be robust with respect to an electric field induced gap in the bulk of the flake. Recent theoretical investigations (*5, 6*) have predicted the presence of such edge states for arbitrary edge reconstructions (armchair edges excluded). These edge states would be able to mix and open a gap at the edge of the sample that is smaller than the bulk gap, possibly of the size of the measured activation gap in **Figure S1**.

### 3. Mesoscopic conductance fluctuations at B=0

At small densities and nonzero electric fields we observe fluctuations of the conductance as function of electric field and density (**Figure S2**). The typical periodicity of these fluctuations is *8 mV/nm* in the electric field axis and *2x10$^9$ 1/cm$^2$* in the density axis. The bias-dependence of the conductance as function of electric field at constant density is shown in **Figure S2**. The conductance maps resemble those of Coulomb charging with typical charging energies of about 1 mV. Furthermore, signs of excited states similar to those observed in etched monolayer quantum dots (*7*) can be seen, as marked by the arrows in **Figure S2**.

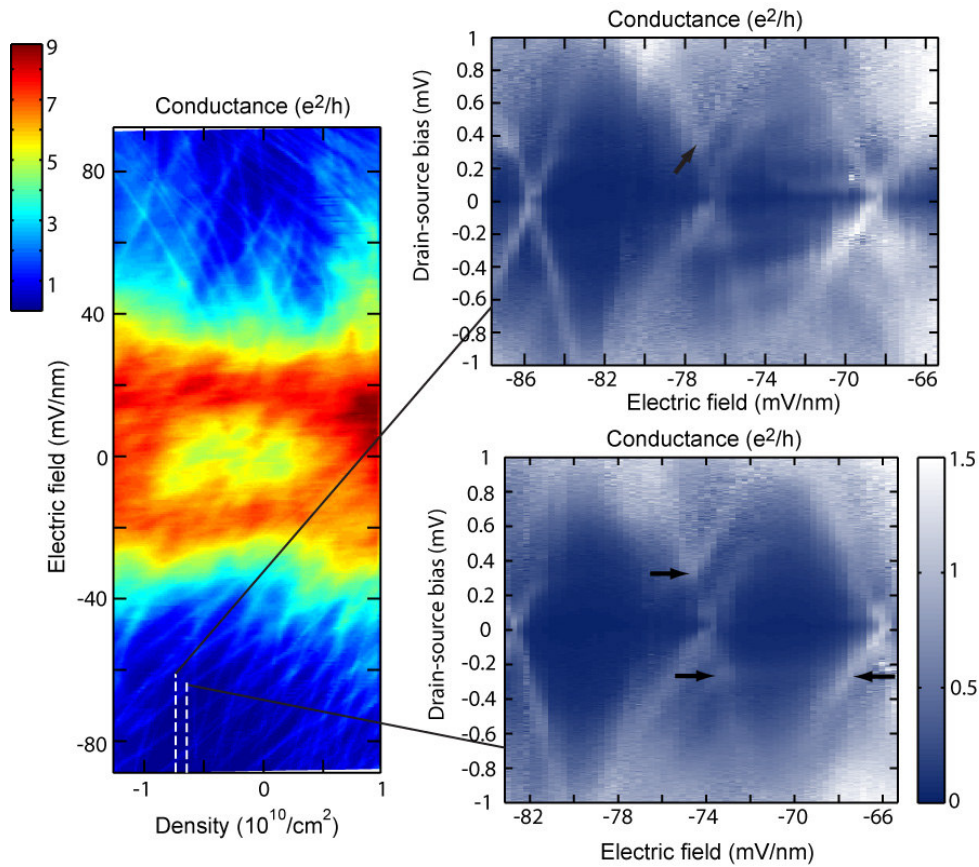

**Figure S2:** Conductance as function of applied electric field and density at zero magnetic field and 60 mK. Regions of increased conductance can be clearly identified to have a dependence of both the electric field and density. Two insets show the bias dependence of the conductance as function of electric field at slightly different densities.

## 4. Model showing the relation between the suggested LL diagram and the observed transport features

A simple model of a linear dispersion of the density of states (DOS) of four Landau levels (LLs) as function of energy and electric field at nonzero constant magnetic field is shown in **Figure S3a**. The broadening of the LLs due to disorder has been accounted for by taking the DOS of the LLs broadened by a Lorentzian function. In **Figure S3b** the charge carrier density as function of electric field has been obtained by numerically integrating over the energy dispersion at constant electric field. In **Figure S3c** the dispersion of the LLs is shown as function of electric field and the carrier density taken from **Figure S3b**. Four regions of increased DOS can be seen that correspond to the crossing of LLs at which the charge carrier density is increased as well. In transport experiments, points of increased charge carrier density correspond to regions of increased conductance, and makes this simple model consistent with our transport experiments.

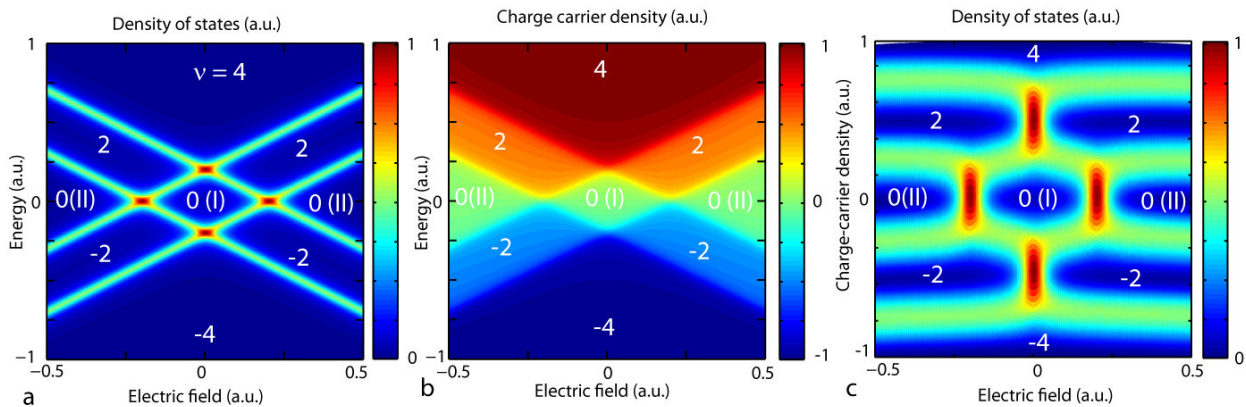

**Figure S3:** (a) Dispersion of the DOS of LLs as function of electric field and energy. (b) Charge carrier density as function of energy and electric field. (c) Dispersion of the DOS of LLs as function of charge carrier density and electric field. The blue regions correspond to regions of constant filling factor.

## 5. Map of conductance versus density at large magnetic field showing the observation of ν = -1, -2, -3 at zero electric field as well as the better development of ν= -3

According to Barlas et al. (*8*) even at zero applied electric field all broken symmetry states should be visible. In our measurements this can be seen at for example a magnetic field of 7.8 T as detailed in **Figure S4**. Here, also ν = -3 is seen to be stronger, even though it is not fully quantized. We have subtracted a resistance of 5 kΩ to account for contact resistance.

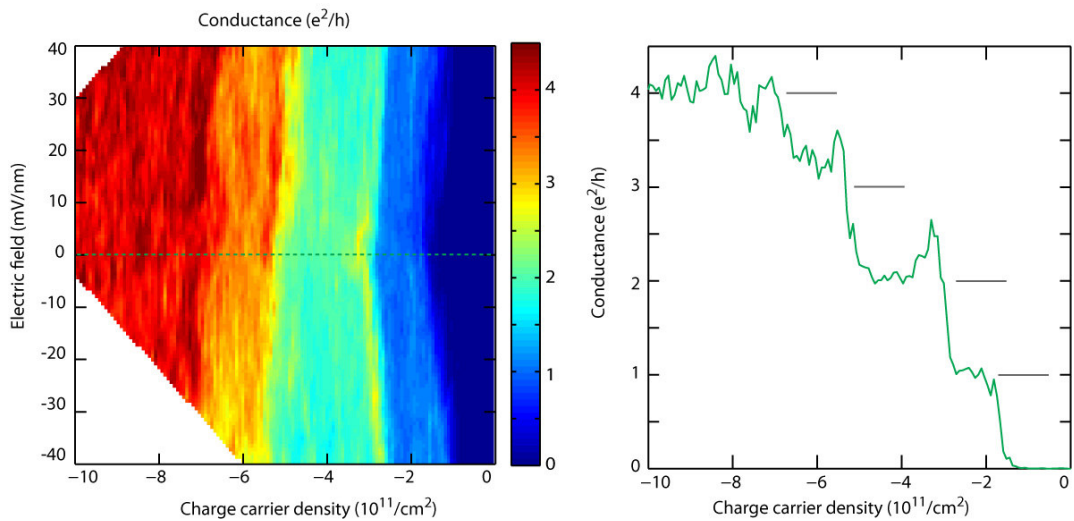

**Figure S4 (a)** Map of the conductance at B = 7.8 T and 60 mK. **(b)** Cut of the conductance as function of density at zero applied electric field.

### 6. Temperature dependence of the transition region at B = 4.5 T ; E = -65 mV/nm

At large electric and magnetic fields the region of increased conductance at the crossing point between the two ν = 0 phases exibits a temperature dependene as shown in **Figure S5**. In the case that disorder is present, it is natural to assume that LLs mix leading to the opening of a gap consistent with our observed temperature dependence.

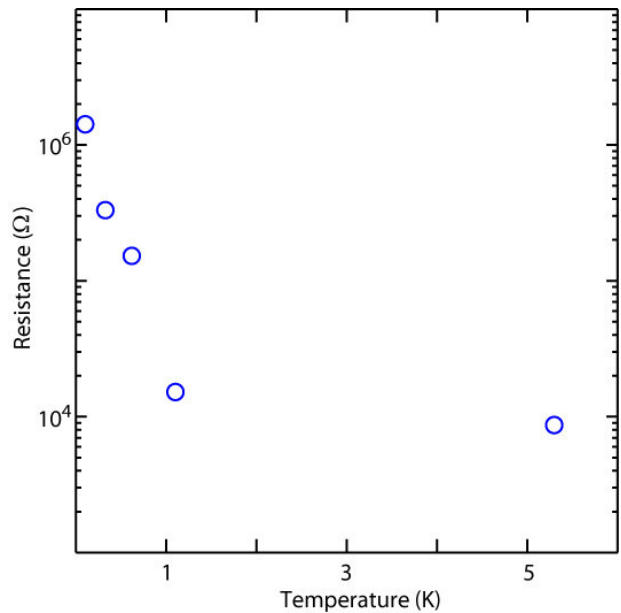

**Figure S5:** Temperature dependence of the resistance at the transition point between the two ν = 0 phases.